\documentclass[aps,11pt,prd,groupedaddress,nofootinbib,notitlepage,eqsecnum,preprintnumbers]{revtex4-2}
\usepackage[utf8]{inputenc}
\usepackage{hyperref}
\usepackage{xcolor}
\usepackage{graphicx}
\usepackage{amsmath,amssymb}
\usepackage{bm}
\usepackage{comment}
\usepackage[shortlabels]{enumitem}
\usepackage{overpic}
\usepackage{ulem}
\usepackage{makecell}
\usepackage{tabularx}
\usepackage{booktabs}
\usepackage{diagbox}
\usepackage{pifont}
\usepackage{etoolbox} 

\makeatletter
\newcommand{\customlabel}[2]{%
   \protected@write \@auxout {}{\string \newlabel {#1}{{#2}{\thepage}{#2}{#1}{}} }%
   \hypertarget{#1}{#2}
}
\makeatother

\def\calR{{\cal R}}

\begin{document}
\title{Probing Primordial Black Hole Scenarios \\ with Terrestrial Gravitational Wave Detectors}

\author{\textsc{Guillem Dom\`enech$^{a,b}$}}
    \email{{guillem.domenech}@{itp.uni-hannover.de}}
\author{\textsc{Misao Sasaki$^{c,d,e}$}}
\email{{misao.sasaki}@{ipmu.jp}}

\affiliation{$^a$Institute for Theoretical Physics, Leibniz University Hannover, Appelstraße 2, 30167 Hannover, Germany.}
\affiliation{$^b$ Max-Planck-Institut für Gravitationsphysik, Albert-Einstein-Institut, 30167 Hannover, Germany}
\affiliation{$^c$ Kavli Institute for the Physics and Mathematics of the Universe (WPI), The University of Tokyo Institutes for Advanced Study, The University of Tokyo, Chiba 277-8583, Japan}
\affiliation{$^d$ Center for Gravitational Physics and Quantum Information, Yukawa Institute for Theoretical Physics, Kyoto University, Kyoto 606-8502, Japan} 
\affiliation{$^e$ Leung Center for Cosmology and Particle Astrophysics, National Taiwan University, Taipei 10617, Taiwan}

\begin{abstract}
It is possible that primordial black holes consitute (or consituted) a significant fraction of the energy budget of our universe. Terrestrial gravitational wave detectors offer the opportunity to test the existence of primordial black holes in two different mass ranges, from $10^2\,{\rm g}-10^{16}\,{\rm g}$ to $10^{-6}\,M_\odot-100 \,M_\odot$. The first mass window is open via induced gravitational waves and the second one by gravitational waves from binary mergers. In this review, we outline and explain the different gravitational wave signatures of primordial black holes that may be probed by terrestrial gravitational wave detectors, such as the current LIGO/Virgo/KAGRA and future ones like Einstein Telescope and Cosmic Explorer. We provide rough estimates for the frequency and amplitude of the associated GW background signals. We also discuss complementary probes for these primordial black hole mass ranges. 
\end{abstract}

\preprint{YITP-24-04}

\maketitle

\section{Introduction \label{sec:intro}}

Our universe could contain more black holes than those expected from the result of stellar evolution. Such non-astrophysical black holes, so-called Primordial Black Holes (PBHs\footnote{This acronym was first used in Ref.~\cite{Carr:1975qj} in 1975.}), could have formed in the very early universe by an event of cosmological proportions. After formation, a PBH decays via Hawking radiation \cite{Hawking:1974rv,Hawking:1975vcx}, with a lifetime proportional to the cube of the initial mass. A crude estimate then tells us that PBHs heavier than $10^{15}\,{\rm g}$ are long-lived and still with us today (and so have a lifetime larger than the age of the universe), while lighter PBHs have already evaporated.\footnote{As a curiosity, we note that if not for quantum physics (that is Hawking radiation), very light PBHs could easily eventually dominate the universe.} See, \textit{e.g.}, Refs.~\cite{Khlopov:2008qy,Sasaki:2018dmp,Carr:2020gox,Green:2020jor,Escriva:2022duf,Ozsoy:2023ryl} for recent reviews on PBHs.

As surprising as it may seem, PBHs could be (or could have been) an essential component of our universe, according to current observations \cite{Carr:2020gox,Carr:2023tpt}. On one hand, long-lived PBHs could explain the dark matter \cite{Bellomo:2017zsr,Carr:2017jsz,Inomata:2017okj,Bartolo:2018rku,Bartolo:2018evs,Carr:2020xqk,Chakraborty:2022mwu}, the microlensing events reported by HSC \cite{Niikura:2017zjd} and OGLE \cite{Mroz:2017mvf,Niikura:2019kqi,Gouttenoire:2023pxh},  some of the LIGO/Virgo/KAGRA (LVK) black hole binary mergers \cite{Bird:2016dcv,Sasaki:2016jop,Wong:2020yig,Hutsi:2020sol,Franciolini:2021tla} and the seeds of supermassive black holes \cite{Kawasaki:2012kn,Bernal:2017nec,Carr:2018rid,Lu:2023xoi,Silk:2024rsf}. On the other hand, evaporated PBHs could have totally reheated the universe \cite{Carr:1976zz,Chapline:1976au,Lidsey:2001nj,Hidalgo:2011fj,Anantua:2008am} and played an important role in generating the baryon asymmetry of the universe \cite{Turner:1979bt,Turner:1979zj,Barrow:1990he,Dolgov:2000ht,Baumann:2007yr,Fujita:2014hha,Morrison:2018xla,Smyth:2021lkn,Garcia-Bellido:2019vlf,Datta:2020bht,Perez-Gonzalez:2020vnz,Bernal:2022pue,Cheek:2022mmy,He:2022wwy,Gehrman:2022imk,Ghoshal:2023fno,Altavista:2023zhw} and/or particle dark matter \cite{Samanta:2021mdm,Cheek:2021cfe,Cheek:2021odj,Kim:2023ixo}.\footnote{PBH remnants after complete evaporation, if any, could also play the role of dark matter \cite{MacGibbon:1987my,Alexander:2007gj}.
See also \cite{Domenech:2023mqk}.} We provide some numbers on the corresponding PBH mass windows to the scenarios discussed above in Tab.~\ref{tab:table0}.

\begin{table}
\def\arraystretch{1.75}
\setlength\tabcolsep{1.5mm}
\begin{tabular}{|c|c|c|c|c|c|}
\toprule[2.0pt]\addlinespace[0mm]
\makecell{\textbf{PBH}\\ \textbf{scenario}} & \makecell{\textbf{Reheating}  }  & \makecell{\textbf{Dark matter}   } & \makecell{\textbf{HSC \& OGLE}\\ \textbf{events}    } & \makecell{\textbf{LVK}\\\textbf{black holes}  }& \makecell{\textbf{SMBH}\\\textbf{seeds}  }\\
\hline
\makecell{\makecell{\textbf{Mass window}   }} &$1-10^{9}\,{\rm g}$& $10^{17}-10^{24}\,{\rm g}$& $10^{-8}M_\odot$ \& $10^{-4}M_\odot$ & $1\,-10^{2}\,M_\odot$& $10^{2}-10^{6}\,M_\odot$\\
\toprule[2.0pt]
\end{tabular}
\caption{Summary of interesting mass ranges for various PBH scenarios, such as PBH reheating, PBH dark matter, HSC and OGLE microlensing events, LVK merger events and seeds of SuperMassive Black Holes (SMBHs). The mass range in the PBH reheating scenario follows from viability: $1\,{\rm g}$ PBHs form right after GUT scale inflation and $10^9\,{\rm g}$ PBHs reheat the universe too close to nucleosynthesis. Note that under the null hypothesis of no PBHs, HSC \cite{Niikura:2017zjd} and OGLE \cite{Niikura:2019kqi} respectively probe the mass ranges $10^{-11}-10^{-6}\,M_\odot$ and $10^{-6}-10^{-3}\,M_\odot$. Most interestingly, HSC found one candidate event with a mass of about $10^{-8}M_\odot$ and OGLE found 6 candidates with masses around  $10^{-4}M_\odot$. A solar mass is $M_\odot\approx 2\times 10^{33}\,{\rm g}$. \label{tab:table0}}
\end{table}

Finding evidence for PBHs would indicate new physics beyond the standard model of cosmology and/or particle physics. For instance, cosmic events that may result in PBHs include the collapse of large primordial fluctuations from cosmic inflation \cite{Hawking:1971ei,Carr:1974nx,Passaglia:2021jla} (see Ref.~\cite{Ozsoy:2023ryl} for a review), phase transitions \cite{Crawford:1982yz,Kodama:1982sf,Garriga:2015fdk,Deng:2016vzb,Liu:2021svg,He:2022amv,Kawana:2022olo,Huang:2023mwy,Escriva:2023uko}, the collapse of Q-balls \cite{Cotner:2016cvr,Cotner:2019ykd,Flores:2021jas} and fifth forces in the early universe \cite{Amendola:2017xhl,Flores:2020drq,Domenech:2021uyx,Domenech:2023afs}. Interestingly, there are several Gravitational Wave (GW) signals associated with PBHs, which can provide strong evidence for (or rule out) the existence of PBHs.  In addition to GWs from PBH binaries \cite{Nakamura:1997sm,Bird:2016dcv,Sasaki:2016jop,Ali-Haimoud:2017rtz,Raidal:2018bbj,Vaskonen:2019jpv,Wang:2019kaf,Braglia:2021wwa,Atal:2022zux,Franciolini:2021xbq,Franciolini:2022ewd}, there are induced GWs associated with PBH formation \cite{Saito:2008jc,Saito:2009jt,Bugaev:2009zh,Bugaev:2010bb} and PBH reheating \cite{Inomata:2020lmk,Papanikolaou:2020qtd,Domenech:2020ssp,Domenech:2021wkk,Papanikolaou:2021uhe,Papanikolaou:2022chm}, as well as high frequency GWs from Hawking evaporation \cite{Dolgov:2011cq,Dong:2015yjs,Hooper:2019gtx,Masina:2020xhk,Masina:2021zpu,Arbey:2021ysg,Ireland:2023avg}. Except for nearby PBH binaries, all such GW signals contribute to the cosmic GW background.

The focus of this review are GW signals of PBHs accessible to terrestrial GW detectors, which happen to include the PBH reheating scenario and PBHs with masses ranging from an earth mass to hundred solar masses. Interestingly, there are tentative hints of solar and sub-solar mass compact objects in the LVK data \cite{Clesse:2020ghq,Morras:2023jvb,LIGOScientific:2022hai,Prunier:2023cyv} and solar-mass objects in quasar broad emission lines \cite{Hawkins:2023nkq}.\footnote{Solar-mass BHs could also be explained by transmuted BHs from the capture of small PBHs by neutron stars \cite{Takhistov:2020vxs}.} There are also the moon-mass and earth-mass microlensing events respectively reported by HSC \cite{Niikura:2017zjd} and OGLE \cite{Mroz:2017mvf,Niikura:2019kqi}. For simplicity, although our discussion may apply to more general situations,\footnote{In general terms, there are basically two relevant comoving scales in the problem, the Schwarzschild radius of the PBHs (or, equivalently, the Hubble radius at PBH formation) and the mean inter-PBH separation (related to the initial fraction of PBHs) plus redshift effects due to the expansion of the universe. Thus, our order of magnitude estimates should still give a rough idea in more general settings, at least for the frequency of the associated GWs.} we will specialize to the case of PBH formation from large primordial fluctuations. In that case, the planet to solar mass window has direct connections to the recent Pulsar Timing Arrays (PTAs) results on a possible GW background \cite{NANOGrav:2023gor,NANOGrav:2023hde,EPTA:2023fyk,EPTA:2023sfo,EPTA:2023xxk,Reardon:2023gzh,Zic:2023gta,Reardon:2023zen,Xu:2023wog,InternationalPulsarTimingArray:2023mzf}, if interpreted as the induced GW signal \cite{NANOGrav:2023hvm,Dandoy:2023jot,Franciolini:2023pbf,Franciolini:2023wjm,Inomata:2023zup,Cai:2023dls,Wang:2023ost,Liu:2023ymk,Unal:2023srk,Figueroa:2023zhu,Yi:2023mbm,Zhu:2023faa,Firouzjahi:2023lzg,Li:2023qua,You:2023rmn,Balaji:2023ehk,HosseiniMansoori:2023mqh,Zhao:2023joc,Liu:2023pau,Yi:2023tdk,Bhaumik:2023wmw,Choudhury:2023hfm,Yi:2023npi,Harigaya:2023pmw,Basilakos:2023xof,Jin:2023wri,Cannizzaro:2023mgc,Zhang:2023nrs,Liu:2023hpw,Choudhury:2023fwk,Tagliazucchi:2023dai,Basilakos:2023jvp,Inomata:2023drn,Li:2023xtl,Domenech:2023dxx,Gangopadhyay:2023qjr,Cyr:2023pgw,Lozanov:2023rcd} (see also Refs.~\cite{Huang:2023chx,Gouttenoire:2023nzr,Depta:2023qst} for the merger of supermassive PBHs).

This review is organized as follows. In \S~\ref{sec:formation}, we explain the various GW signals associated with PBHs and provide order of magnitude estimates for the frequency and amplitude. In \S~\ref{sec:testPBH}, we focus on the potential of terrestrial GW detectors to probe PBH scenarios. Then, in \S~\ref{sec:complementary} we briefly discuss other complementary probes to terrestrial GW detectors. We conclude our work with further discussions in \S~\ref{sec:conclusions}. Useful references to dig into the details of PBHs from the collapse large primordial fluctuations are Ref.~\cite{Sasaki:2018dmp} for a general review, Ref.~\cite{Carr:2020gox} for an extensive review on available constraints, Ref.~\cite{Escriva:2022duf} for the details of PBH formation and Ref.~\cite{Ozsoy:2023ryl} for the connection to inflation.

\section{GW signals of black holes from the early universe \label{sec:formation}}

Suppose that in the early, radiation dominated universe there is a large positive fluctuation in the spatial curvature of the universe on scales larger than the Hubble radius (sometimes also called Hubble horizon). If such a positive curvature fluctuation is as large as the square of the Hubble parameter when its comoving scale becomes comparable to the Hubble radius, the Hubble volume of that region behaves like a closed universe. And, a closed universe eventually collapses on itself. But, seen from outside of that particular Hubble volume, it is the formation of a PBH. 
For more details on this nice geometrical picture see Ref.~\cite{Sasaki:2018dmp}. Consequently the mass of the resulting PBH will be proportional to that contained inside the Hubble radius at formation, say $H_{\rm f}^{-1}$ where $H_{\rm f}$ is the expansion rate of the universe at formation. Explicitly, we have that
\begin{align}\label{eq:MPBHf}
M_{\rm PBH,f}=4\pi\gamma\frac{M_{\rm pl}^2}{H_{\rm f}}&\approx 10^{-4} M_\odot\left(\frac{T_{\rm f}}{10 \, {\rm GeV}}\right)^{-2}\left(\frac{g_\rho(T_{\rm f})}{106.75}\right)^{-1/2}\,,
\end{align}
where $\gamma$ is the efficiency factor usually taken to be $\gamma\sim 0.2$ \cite{Sasaki:2018dmp} and in the last step we wrote the Hubble radius in terms of the temperature $T_{\rm f}$ of the radiation filling the early universe at formation. $g_\rho(T_{\rm f})$ is the effective number of massless degrees of freedom in the energy density of the primordial plasma. Whenever needed we assume the standard model of particles and use the values of Ref.~\cite{Saikawa:2018rcs}. As a curiosity, note that the Schwarschild radius of the PBH is of the order of the Hubble radius at formation, namely $r_{\rm PBH,f}=2GM_{\rm PBH,f}={\gamma}/{H_{\rm f}}$. The Hubble radius $H_{\rm f}$, or the temperature $T_{\rm f}$, is also related to the comoving scale $k_{\rm f}$ of the relevant fluctuation, which in turn is related to the time of generation during inflation and frequency of the induced GWs. We provide an estimate below.

Superhorizon curvature fluctuations are, in fact, the initial conditions of the standard cosmological model. They explain the Cosmic Microwave Background (CMB) anisotropies and later give rise to galaxies and other structures we see today in our universe. CMB observations also provide strong evidence that such primordial fluctuations were generated during a period of cosmic inflation \cite{Akrami:2018odb} (see, \textit{e.g.}, \cite{Guzzetti:2016mkm,Achucarro:2022qrl} for recent brief reviews focused on GWs and other future prospects). The general and simplest prediction of inflation is that of random Gaussian superhorizon curvature fluctuations with an almost scale invariant spectrum, which actually arise from vacuum quantum fluctuations \cite{Mukhanov:1985rz,Sasaki:1986hm,
Mukhanov:1990me}. So, strictly speaking, the universe is filled with fluctuations of all sizes and amplitudes. The issue is that the amplitude of the primordial spectrum measured by the CMB is about $10^{-9}$ \cite{Planck:2018vyg} and the probability that one fluctuation is large enough to form a PBH is absurdly exponentially suppressed. But, new physics during inflation can enhance the spectrum of fluctuations up to an amplitude of $10^{-2}$, yielding an interesting fraction of PBHs in the universe.\footnote{See Refs.~\cite{Kristiano:2022maq,Riotto:2023hoz,Kristiano:2023scm,Riotto:2023gpm,Firouzjahi:2023aum,Firouzjahi:2023ahg,Franciolini:2023lgy,Tasinato:2023ukp,Cheng:2023ikq,Fumagalli:2023hpa,Inomata:2022yte,Fumagalli:2023loc,Iacconi:2023ggt,Davies:2023hhn} for an interesting ongoing discussion on the impact of one loop quantum effects.} Unfortunately, we will not dwell into the physics during inflation, which would deserve a whole new chapter. Instead we refer the reader to Ref.~\cite{Ozsoy:2023ryl} (and references therein) for a review of inflation focused on PBHs.

Although there has been substantial progress in recent years in the understanding of when and how PBHs form, the precise condition for the PBH formation is still under study (see e.g.,~\cite{Escriva:2021aeh, Yoo:2022mzl,Germani:2023ojx}). Since the PBH formation is not a main topic of this review, for the sake of simplicity,
here we consider the PBH formation in the naive Press-Schechter formalism \cite{Press:1973iz}.
In this formalism, we assume that fractional density fluctuations, $\delta\equiv\delta\rho/\rho$, above a certain threshold, $\delta_{\rm th}$, form PBHs. The abundance of PBHs is then given by the probability to have $\delta$ above such a threshold \cite{Sasaki:2018dmp}, namely,
\begin{align}\label{eq:betaPS}
\beta=\gamma \int_{\delta_{\rm th}}^1 \frac{d\delta}{\sqrt{2\pi}\sigma_{M_{\rm PBH,f}}}e^{-\frac{\delta^2}{2\sigma_{M_{\rm PBH,f}}^2}}\approx\gamma\frac{\delta_{\rm th}}{\sqrt{2\pi}\sigma_{M_{\rm PBH,f}}}e^{-\frac{\delta^2}{2\sigma_{M_{\rm PBH,f}}^2}} \,,
\end{align}
where $\gamma$ is the efficiency factor, $\sigma_{M_{\rm PBH,f}}$ is the variance of fluctuations smoothed over the comoving scale $R_{\rm f}$ (basically $1/k_{\rm f}$), and in the last step we assumed that $\sigma_{M_{\rm PBH,f}}\ll \delta_{\rm th}$. Note that we assumed that the statistics of the primordial fluctuations are Gaussian.
The smoothed variance of fluctuations reads
\begin{align}\label{eq:varianceMPBHf}
\sigma_{M_{\rm PBH,f}}^2=\int_{-\infty}^\infty d\ln k \,{\cal P}_\delta(k)\, W^2(kR_{\rm f})=\frac{16}{81}\int_{-\infty}^\infty d\ln k\,{\cal P}_{\calR}(k)\,(kR_{\rm f})^4\, W^2(kR_{\rm f})\,.
\end{align}
In Eq.~\eqref{eq:varianceMPBHf}, $W(kR_{\rm f})$ is the window function, ${\cal P}_{\calR}$ the primordial spectrum of curvature fluctuations $\calR$ and in the last step we use the relation between density fluctuations and curvature fluctuations on superhorizon scales in a radiation dominated universe \cite{Sasaki:2018dmp}. The smoothing of the fluctuations is necessary to go from the Fourier space results to real space, focusing only on the relevant scale. From Eqs.~\eqref{eq:betaPS} and \eqref{eq:varianceMPBHf} it is clear that the initail fraction of PBHs, $\beta$, is exponentially sensitive to the primordial spectrum ${\cal P}_{\cal R}$. For long-lived PBHs, we may relate the initial fraction of PBHs with the fraction of dark matter in the form of PBHs, ${\rm f}_{\rm PBH}$. Namely,
\begin{align}\label{eq:betafPBH}
\beta&
\approx 3\times 10^{-9}\,{\rm f}_{\rm PBH}\left(\frac{M_{\rm PBH,f}}{M_\odot}\right)^{1/2}\left(\frac{g_\rho(T_{\rm f})}{10.75}\right)^{-3/4}\left(\frac{g_s(T_{\rm f})}{10.75}\right)\,,
\end{align} 
where $g_s(T_{\rm f})$ are the effective degrees of freedom in the entropy. Note that the initial fraction $\beta$ may be initially very small but the fraction of PBHs in the universe grows in time, since the energy density of radiation dilutes faster than that of PBHs. For evaporated PBHs we shall explore their cosmology separately later.

As a word of caution, we would like to stress again that the Press-Schechter formalism is the simplest estimate and we believe that, in view of the current state-of-the-art, it should not be used to accurately predict the fraction of PBHs. We refer the reader to Refs.~\cite{Escriva:2021aeh,Yoo:2022mzl,Germani:2023ojx} for reviews on recent advancements in PBH formation criteria. 
Here we solely mention that the fraction of PBHs is highly dependent on the precise value of the threshold  (which also depends on the equation state of the universe \cite{Musco:2012au,Harada:2013epa} and the radial profile of the fluctuation \cite{Germani:2018jgr} but a typical value is $\delta_{\rm th}\sim 0.45$), the tail of the probability distribution of curvature fluctuations, the critical collapse, the detailed formalism and the window function one chooses. For example, non-Gaussianity of primordial fluctuations drastically changes the amount of PBHs produced \cite{DeLuca:2019qsy,Kehagias:2019eil,Kitajima:2021fpq,Young:2022phe,Pi:2022ysn}. This effect is particularly important in the induced GW interpretation of the PTA data \cite{Franciolini:2023pbf,Figueroa:2023zhu,Liu:2023ymk,Li:2023qua}, as assuming Gaussian fluctuations tends to predict too many PBHs (though we stress that there are still uncertainties in the calculations of the exact PBH fraction). Fortunately, as we show later, the GW signals are not sensitive to the non-linear physics of PBH formation. For earlier works on the impact of local non-Gaussianities on the induced GWs see Refs.~\cite{Cai:2018dig,Unal:2018yaa,Atal:2021jyo,Adshead:2021hnm,Abe:2022xur,Yuan:2023ofl,Li:2023xtl}. See also Ref.~\cite{Inui:2023qsd} for an analysis of LVK data including non-Gaussianities in the induced GWs.

\subsection{GW signatures of PBHs \label{subsec:GWsignatures}}

We proceed to describe the GWs associated with PBHs and provide estimates for their frequency and amplitude. Before doing that though, we would like to note that there are other reviews on GW signatures of PBHs. For instance, there is an extensive review by the LISA cosmology working group \cite{LISACosmologyWorkingGroup:2023njw}, with focus on LISA capabilities to test PBH scenarios (see also Ref.~\cite{Franciolini:2021nvv}). And a pedagogical and intuitive introduction can be found in Ref.~\cite{Domenech:2023fuz}. Here we aim for a complementary and concise summary focused on terrestrial GW detectors, such as LVK, Einstein Telescope and Cosmic Explorer. For simplicity and analytical viability, we assume a monochromatic (or almost monochromatic) primordial spectrum of curvature fluctuations and PBH mass function. When pertinent, we also discuss and cite works on broad mass functions. We also assume the standard model of particle physics and take appropriate values for the effective degrees of freedom when necessary. To recover the dependence on the effective degrees of freedom we will refer the reader to the relevant references.

We classify the GW signals associated with PBHs into:
\begin{itemize}
\item[\textit{(i)}] GWs associated with PBH formation,
\item[\textit{(ii)}]  GWs associated with PBH reheating, and,
\item[\textit{(iii)}] GWs from PBH binary mergers.
\end{itemize}
The first two signals are induced GWs, while the third are typical GWs from black hole binaries (either resolved or unresolved). At this point, it is important to clarify that, while GWs from PBH binaries (and Hawking evaporation) come from PBHs themselves, that is not the case for induced GWs. Induced GWs \cite{Tomita,Matarrese:1992rp,Matarrese:1993zf,Ananda:2006af,Baumann:2007zm} (see Refs.~\cite{Yuan:2021qgz,Domenech:2021ztg,Domenech:2023jve} for recent reviews), sometimes also called secondary GWs, are a consequence of the cosmological process that led to PBH formation (if associated with PBH formation) or the cosmological process that resulted from the complete PBH evaporation (if associated with PBH reheating). A more concrete explanation for the GWs associated with PBH formation is as follows. In order to form PBHs there must be highly inhomogeneous concentrations of matter involved in the early universe, resulting in density waves and large anisotropic stresses. The latter are responsible for the generation of induced GWs. A similar logic applies after PBH reheating. 

Before proceeding with some estimates for the frequency and amplitude of the associated GWs, let us clarify the notation in what follows. When dealing with induced GWs, we provide the amplitude of the GW spectrum evaluated at a time when a given GW frequency is sufficiently inside the horizon to be regarded as a proper wave and, therefore, as a radiation fluid. After that epoch the energy density ratio of GWs, defined by $\Omega_{\rm GW}=\rho_{\rm GW}/\rho_{\rm total}$ where $\rho$ means energy density (see, \textit{e.g.}, Ref.~\cite{Maggiore:1900zz} for the definition of the energy density of GWs in cosmology), is mostly constant in the early universe; it only changes when there is a change in the effective degrees in energy density and entropy. 
When needed, we take this time to be the epoch of horizon crossing (given by $k=aH$ where $k$ is the comoving wavenumber and $a$ the scale factor), although more realistic estimates suggest it may be at least $2$ e-foldings later \cite{Smith:2006nka,Pritchard:2004qp}. We note however that this does not matter much for the GWs we consider, since in the very early universe the number of the effective degrees of freedom is basically constant. 
The spectral density of GWs today can then be written as \cite{Inomata:2018epa,Domenech:2021ztg}
\begin{align}\label{eq:GWstoday}
\Omega_{\rm GW,0}h^2&=1.62\times 10^{-5}\left(\frac{\Omega_{\rm rad,0}h^2}{4.18\times 10^{-5}}\right)\left(\frac{g_{\rho}(T_*)}{106.75}\right)\left(\frac{g_{s}(T_*)}{106.75}\right)^{-4/3}\Omega_{\rm GW,*}\,,
\end{align}
where $\Omega_{\rm GW,*}$ is the spectral density of GWs well enough inside the Hubble horizon and $\Omega_{\rm rad,0}h^2$ is the energy density fraction of radiation today given by Planck \cite{Planck:2018vyg}. Note that, for convenience, we drop the subscript $*$ in the estimates below. We provide a summary of our estimates for the frequency and amplitude evaluated today via Eq.~\eqref{eq:GWstoday} in Tab.~\ref{tab:table1}.
\\

\begin{table}
\def\arraystretch{2.25}
\setlength\tabcolsep{1.9mm}
\begin{tabular}{|c|c|c|}
\toprule[2.0pt]\addlinespace[0mm]
\makecell{\textbf{GW background} \\\textit{associated with} PBH} & \makecell{\textbf{Peak frequency}\\$(f_{\rm peak}[{\rm kHz}])$ }& \makecell{\textbf{Peak amplitude and spectral index today}\\ $(\Omega_{\rm GW,0}h^2\approx\Omega^{\rm peak}_{\rm GW,0}h^2\times(f/f_{\rm peak})^\alpha)$ }  \\
\toprule\addlinespace[0mm]
\makecell{\textbf{Formation}$^*$   \\ {(Adiabatic iGWs$^\dag$)}} & $12\left(\frac{M_{\rm PBH,f}}{10^8\rm g}\right)^{-\tfrac{1}{2}}$
& {$ 10^{-5} {\cal A}_{\calR}^2 \,\,;\,\, \alpha\approx 3\,(2)$} \\
\hline
\makecell{\textbf{Reheating I}$^{**}$   \\{(Isocurvature iGWs)}} & $1.7\,\left(\frac{M_{\rm PBH,f}}{10^4\,{\rm g}}\right)^{-\tfrac{5}{6}}$
& $2\times10^{-7}\left(\frac{\beta}{10^{-6}}\right)^{\tfrac{16}{3}}\left(\frac{M_{\rm PBH,f}}{10^4\,{\rm g}}\right)^{\tfrac{34}{9}}\,\,;\,\, \alpha\approx \frac{11}{3}$  \\
\hline
\makecell{\textbf{Reheating II}$^{**}$  \\{(Adiabatic iGWs)}} & $2.5\,\left(\frac{\Omega_{\rm GW,0}h^2}{10^{-15}}\right)^{\tfrac{3}{7}}\left(\frac{M_{\rm PBH,f}}{10^2\,{\rm g}}\right)^{-\tfrac{3}{2}}$
& $ 10^{-15}\left(\frac{\beta M_{\rm PBH,f}}{10^{-3}\,{\rm g}}\right)^{-\tfrac{14}{9}}
   e^{-\left(\frac{\beta M_{\rm PBH,f}}{10^{-3}\,{\rm g}}\right)^{-{4}/{3}}}\,\,;\,\, \alpha\approx 7$ \\
\hline
\makecell{\textbf{Mergers}  \\{(Unresolved binaries)} }& $ 4.4\left(\frac{M_{\rm PBH,f}}{M_\odot}\right)^{-1}$
& $2\times 10^{-8}\left(\frac{M_{\rm PBH,f}}{M_\odot}\right)^{\tfrac{5}{37}}\left(\frac{{\rm f}_{\rm PBH}}{0.01}\right)^{\tfrac{53}{37}}\,\,;\,\,\alpha\approx \frac{2}{3}$ \\
\toprule[2.0pt]
\end{tabular}
\flushleft{$^*$ GWs from the cosmological process leading to PBH formation, not from formation of individual PBHs.\newline
$^{**}$ GWs from the cosmic dynamics after full PBH evaporation, not from Hawking evaporation.\newline 
$^\dag$ iGWs: induced GWs.}
\caption{Summary of estimates for the frequency and amplitude of GW signatures associated with PBH discussed in points \ref{i}, \ref{ii}, \ref{iii}. ${\cal A}_{\calR}$ is the amplitude of the primordial spectrum of curvature fluctuations, see Eq.~\eqref{eq:lognormalpeak}. Note that all GW signals have a sharp cut-off for $f>f_{\rm peak}$, since we assume a monochromatic PBH mass function. The power-law index for $f<f_{\rm peak}$ is given by the value of $\alpha$. $\Omega^{\rm peak}_{\rm GW,0}h^2$ is the peak frequency of the spectral density of GWs evaluated today using  Eq.~\eqref{eq:GWstoday}. \label{tab:table1}}
\end{table}

\noindent\textbf{\customlabel{i}{\textit{(i)}}  GWs associated with PBH formation.} Large enough and rare primordial fluctuations collapse to form PBHs. The other, not large enough, average\footnote{By average we have in mind the root mean square of primordial fluctuations.} primordial curvature fluctuations generate density waves with amplitudes damped in time by the primordial plasma. These are the main source of induced GWs.\footnote{There are also scalar-tensor induced GWs \cite{Chang:2022vlv,Yu:2023lmo,Bari:2023rcw,Picard:2023sbz} which are often subdominant but might have distinctive features \cite{Bari:2023rcw}.} Thus, the largest production of induced GWs occurs at around the time of Hubble horizon crossing of the typical primordial fluctuation, say with comoving wavenumber $k_{\rm p}$. For an almost monochromatic primordial spectrum, $k_{\rm p}$ determines the frequency at which the spectrum of the GW background peaks, as well as the typical mass of the PBHs by Eq.~\eqref{eq:MPBHf} (and using that $k_{\rm f}=k_{\rm p}$). The frequency evaluated today is given by\footnote{A more detailed formula including the effective degrees of freedom in energy density and entropy is given by
\begin{align}
f_{\rm formation}=\frac{k_{\rm p}}{2\pi a_0}\approx 1.2\times 10^{4}\,{\rm Hz}\left(\frac{M_{\rm PBH,f}}{10^8\,\rm g}\right)^{-1/2}\left(\frac{g_\rho(T_{\rm f})}{106.75}\right)^{1/4}\left(\frac{ g_{s}(T_{\rm f}) }{106.75}\right)^{-1/3}\,.
\end{align}
The prefactors are important for PBH masses above $10^{-3}M_\odot$ as the corresponding temperature is below $1\,{\rm GeV}$. However, this only concerns $\mu$Hz-nHz frequencies and, therefore, it is not relevant for this review.}
\begin{align}\label{eq:ffomration}
f_{\rm formation}=\frac{k_{\rm p}}{2\pi a_0}\approx 12\,{\rm kHz}\left(\frac{M_{\rm PBH,f}}{10^8\, \rm g}\right)^{-1/2}\,,
\end{align}
where we used Eq.~\eqref{eq:MPBHf} to replace $k_{\rm f}$ with $M_{\rm PBH,f}$ since it is more convenient for our discussions.

The amplitude of the GW spectrum depends on the primordial spectrum of curvature fluctuations. For concreteness, we assume a log-normal spectrum given by
\begin{align}\label{eq:lognormalpeak}
\mathcal{P}_{\calR}(k)=\frac{\mathcal{A_{\calR}}}{\sqrt{2\pi}\Delta}\exp\left[-\frac{\ln^2(k/k_{\rm p })}{2\Delta^2}\right],
\end{align}
where $\mathcal{A_{\calR}}$ is the amplitude and $\Delta$ is the logarithmic width of the spectrum. One recovers the Dirac delta case in the limit when $\Delta\to 0$. The log-normal spectrum allows for nice analytical approximations for the induced GW spectrum, as derived in Ref.~\cite{Pi:2020otn}. For our purposes though, a good enough estimate for the amplitude at the peak is given by
\begin{align}\label{eq:estimatepeak}
  \Omega^{\rm peak}_{\rm GW,formation} \approx {\cal O}(1-10) {\cal A}_{\calR}^2\,,
\end{align}
where the factor ${\cal O}(10)$ corresponds to sharp primordial spectrum with $\Delta<0.1$. We note, however, that the estimate for the peak amplitude \eqref{eq:estimatepeak} is also valid in more general situations. For instance, for a scale invariant primordial spectrum, the GW spectrum is also scale invariant with amplitude $\sim 0.8{\cal A}_{\calR}^2$. The main difference is the spectral shape. For a sharp peak, there is a sharp cut-off at $f\sim 2f_{\rm formation}$ (above which there are no more GWs produced by momentum conservation) and a low frequency tail going as $\Omega_{\rm GW}\propto f^{2}$ which transitions to $\Omega_{\rm GW}\propto f^{3}$ when far enough from the peak, roughly for $f< 2\Delta \times f_{\rm formation}$ \cite{Pi:2020otn}. It is interesting to note that the $f^{3}$ is a universal infra-red scaling for GWs from localized sources \cite{Cai:2019cdl} but induced GWs in radiation domination contain a logarithmic correction \cite{Cai:2018dig,Yuan:2019wwo}. We also note that these estimates depend on the expansion history of the universe, \textit{e.g.}, on the equation of state of the early universe \cite{Domenech:2019quo,Domenech:2020kqm}. We show the shape of the induced GW spectrum for a log-normal \eqref{eq:lognormalpeak} with $\Delta=0.1$ in blue in Fig.~\ref{Fig:plot}. For general semi-analytical formulas of induced GWs see Refs.~\cite{Espinosa:2018eve,Kohri:2018awv,Domenech:2019quo,Domenech:2020kqm,Domenech:2021ztg}. \\

\noindent\textbf{\customlabel{ii}{\textit{(ii)}}  GWs associated with PBH reheating.} In two nice papers, Inomata et al. \cite{Inomata:2019ivs,Inomata:2019zqy} studied the induced GWs generated during a transition from pressure-less matter domination to radiation domination, having in mind some models of reheating. They found that sudden transitions enhance the production of induced GWs \cite{Inomata:2019ivs}. The main reason is what some of the authors later called the “poltergeist mechanism” \cite{Inomata:2020lmk} (see also Ref.~\cite{Harigaya:2023mhl} for an application using axions). Essentially what happens is the following: Density fluctuations grow during the matter dominated era and then, suddenly, everything (including those large density fluctuations) are converted into radiation. And, radiation wants to propagate. This creates big sound waves and a loud GW signal. It also gives, in general, a peaked GW signal since smaller scale fluctuations have more time to grow. It turns out that PBH evaporation after PBHs dominate the universe is a good example of an almost sudden transition \cite{Inomata:2020lmk}, if one assumes a monochromatic mass function.

PBHs dominate the universe before evaporating if there is a large enough initial fraction of them. This is because their mean energy density redshifts initially as the volume (by the PBH number density conservation), \textit{i.e.} as $a^{-3}$, while the energy density of radiation dominating the universe dilutes as $a^{-4}$. Thus, on one hand, PBHs dominate the universe for $a/a_f>\beta^{-1}$. On the other hand, the time of evaporation, say $t_{\rm eva}$, is solely set by the initial PBH mass. And, if PBHs dominate, this time $t_{\rm eva}$ also determines the Hubble parameter at evaporation, since $H_{\rm eva}\sim 1/t_{\rm eva}$. From that, one determines that the temperature $T_{\rm eva}$ of radiation filling the universe after PBH evaporation is given by
\begin{align}
T_{\rm eva}\approx
30\,{\rm TeV}\,\left(\frac{M_{\rm PBH,f}}{10^4\,{\rm g}}\right)^{-3/2}\,.
\end{align}
Requiring that evaporation occurs much later than domination leads to a lower bound on the initial fraction, namely
\begin{align}
\beta M_{\rm PBH,f}>  6\times 10^{-6}{\rm g}\,.
\end{align}
We can also use that a successful Big Bang Nucleosynthesis (BBN) requires $T_{\rm eva}>4 \,{\rm MeV}$ \cite{Kawasaki:1999na,Kawasaki:2000en,Hannestad:2004px,Hasegawa:2019jsa}, to place an upper bound to the mass given by
\begin{align}\label{eq:boundonmassBBN}
M_{{\rm PBH},\rm f}<5\times 10^8{\rm g}\,.
\end{align} 

There are, at least, two type of sources for induced GWs in the PBH reheating scenario. The first one are PBH number density fluctuations, which was first pointed out by Ref.~\cite{Papanikolaou:2020qtd}. They come from the statistical fluctuations associated with the discreteness of PBHs. And, as PBH appear to a good approximation randomly and uniformly distributed in space, their fluctuations follow a Poisson distribution. The second source are primordial adiabatic curvature fluctuations \cite{Inomata:2020lmk}, although one must extrapolate the results from CMB scales down to very small scales. 

It is important to note that, in both cases, density fluctuations may enter the non-linear regime (\textit{i.e.} $\delta\rho_{\rm PBH}/\rho_{\rm PBH}>1$), due to the growth of fluctuations in a matter dominated universe. Whether one stops the calculations at the onset of the non-linear regime or not, changes the final amplitude of the induced GWs. However, we note that curvature fluctuations, which are the source of induced GWs, remain always within the validity of perturbation theory. In this review we merely follow the approach of used in the corresponding previous works. In the first case, since the PBH density fluctuations are determined by the presence of PBH themselves, we estimate the GWs using the solutions of linear perturbation theory from the results of \cite{Domenech:2020ssp,Domenech:2021wkk}. But, for adiabatic fluctuations, we impose a conservative cut-off in the power spectrum at the comoving scale which becomes non-linear at reheating as in Ref.~\cite{Inomata:2020lmk}, which we call $k_{\text{nl-cut}}$. While the first approach may overestimate the GW signal, the second one underestimates it. A most accurate calculation must deal with the non-linear regime but likely requires the use of numerical simulations. There is also the fact that relaxing the monochromatic assumption for the PBH mass suppresses the induced GW signal \cite{Inomata:2020lmk}, as PBH evaporation becomes more gradual the broader the mass function is.
For recent hybrid N-body and lattice simulations of a gradual transition see Ref.~\cite{Fernandez:2023ddy}. For estimations of the GWs from non-linear structure formation see Ref.~\cite{Dalianis:2020gup} and for possible turbulence after evaporation see Ref.~\cite{Kozaczuk:2021wcl}. See also Ref.~\cite{Flores:2022uzt} for GWs from the structure formation from Yukawa forces in the early universe.\\

\paragraph{GWs associated with PBH reheating I (isocurvature induced GWs):} The type of initial conditions for Poisson PBH density fluctuations are called isocurvature fluctuations \cite{Kodama:1986ud,Malik:2004tf}. This means that the initial PBH density fluctuations are compensated by equal in amplitude but opposite radiation density fluctuations. Basically, since PBHs originate from the collapse of density fluctuations in the radiation, a hole in the original radiation fluid is filled with a PBH. For a review on isocurvature induced GWs see Ref.~\cite{Domenech:2023jve} (see also Refs.~\cite{Domenech:2021and} and \cite{Passaglia:2021jla} for GWs and PBHs from dark matter isocurvature and Refs.~\cite{Lozanov:2023aez,Lozanov:2023knf} for universal (isocurvature) Gravitational Waves associated with solitons). The spectrum of PBH density fluctuations grows as $k^3$ and it is largest at the scale corresponding to the mean inter-PBH separation, below which the PBH gas picture is no longer valid. We call the frequency associated with the mean inter-PBH separation $f_{\rm poisson}$. In terms of the PBH mass this frequency reads \cite{Domenech:2020ssp}
\begin{align}\label{eq:fpossion}
f_{\rm poisson}
\approx 1.7\, {\rm kHz}\,\left(\frac{M_{\rm PBH,f}}{10^4\,{\rm g}}\right)^{-5/6}\,.
\end{align}
Interestingly, the GW peak frequency only depends on the initial PBH mass. The amplitude of GWs at the peak frequency is estimated to be \cite{Domenech:2020ssp}
\begin{align}\label{eq:GWspossion}
  \Omega^{\rm peak}_{\rm GW,poisson} \approx 10^{-2}\left(\frac{\beta}{10^{-6}}\right)^{16/3}\left(\frac{M_{\rm PBH,f}}{10^4\,{\rm g}}\right)^{34/9}\,.
\end{align}
Then, the GW spectral density has a sharp cut-off for $f>f_{\rm possion}$ and goes as $f^{11/3}$ for $f<f_{\rm possion}$. The factors $1/3$ in the exponents come from the fact that curvature fluctuations on small scales are suppressed by the decay in the PBH mass, which goes as $M_{\rm PBH}\approx M_{\rm PBH,f} (1-t/t_{\rm eva})^{1/3}$ \cite{Inomata:2020lmk}. We show the GW spectral density in red in Fig.~\ref{Fig:plot}. From Eq.~\eqref{eq:GWspossion}, we may use current BBN constraints \cite{Cyburt:2004yc,Arbey:2021ysg} (see Ref.~\cite{Grohs:2023voo} for a recent review) to place an upper bound on the initial fraction of evaporated PBHs, which reads
\begin{align}
\beta< 1.5\times10^{-6}\left(\frac{M_{{\rm PBH},\rm f}}{10^4{\rm g}}\right)^{-17/24}\,.
\end{align}
As far as we are aware, this is the only way to put an upper limit to the initial fraction of PBHs. We note that there are also induced GWs produced during the PBH dominated era \cite{Papanikolaou:2020qtd}. But, in the case of a monochromatic PBH mass function, the induced GWs after PBH evaporation constitute the largest contribution to the GW background.\\ 

\paragraph{GWs associated with PBH reheating II (adiabatic induced GWs):} CMB observations measured an almost scale invariant spectrum of curvature fluctuations with amplitude ${{\cal A}^{\rm CMB}_{\calR}}\sim{2\times 10^{-9}}$. If such a spectrum extends to very small scales and PBHs reheat the universe, then it yields an enhanced GW signal. However, if the primordial spectrum extends to arbitrary small scales, density fluctuations on certain scales become non-linear. To avoid such a regime Ref.~\cite{Inomata:2020lmk} imposes a cut-off such that no fluctuation enters the non-linear regime up to evaporation. Borrowing the results from Ref.~\cite{Inomata:2020lmk}, this cut-off is given by
\begin{align}\label{eq:kNL}
k_{\text{nl-cut}}\sim {\cal P}^{-1/4}_{\Phi}(t_{\rm eva})\,k_{\rm eva}\,.
\end{align}
Thus there are only density fluctuations with $k<k_{\text{nl-cut}}$. In Eq.~\eqref{eq:kNL} ${\cal P}_{\Phi}(t_{\rm eva})$ is the spectrum of fluctuations of the gravitational potential $\Phi$ at evaporation and  $k_{\rm eva}=a_{\rm eva}H_{\rm eva}$ is the wavenumber that enters the horizon at evaporation. Translating the cut-off \eqref{eq:kNL} into a frequency evaluated today yields\footnote{To compare with Ref.~\cite{Inomata:2020lmk} one should use that, in their notation, ${\tau_{\rm eq,2}}/{\tau_{\rm eq,1}}\approx 100 \left(\frac{\beta}{10^{-5}}\frac{M_{\rm PBH,f}}{10^2\,{\rm g}}\right)^{2/3}$. We also considered the limit where $k_{\text{nl-cut}}\tau_{\rm eq,1}\gg 1$. And, although the highest GW production happens for $k_{\text{nl-cut}}\tau_{\rm eq,1}\sim 1$, Eq.~\eqref{eq:GWsfnl-cut} still gives a good order of magnitude estimate.}
\begin{align}\label{eq:fnl-cut}
f_{\text{nl-cut}}=\frac{k_{\text{nl-cut}}}{2\pi a_0}
\approx 2.7\, {\rm kHz}\,\left(\frac{\beta}{10^{-5}\,{\rm g}}\right)^{2/3}\left(\frac{M_{\rm PBH,f}}{10^2\,{\rm g}}\right)^{-5/6}
   e^{\left(\frac{{\cal A}^{\rm CMB}_{\calR}}{2\times 10^{-9}}\right)^{-1/2}\left(\frac{\beta}{10^{-5}}\frac{M_{\rm PBH,f}}{10^2\,{\rm g}}\right)^{-4/3}}\,.
\end{align}
In this case the frequency depends on all the model parameters. The GW spectrum peaks at $f_{\text{nl-cut}}$ with amplitude\footnote{In the regime where $k_{\text{nl-cut}}\gg k_{\rm eq}$, with $k_{\rm eq}$ being the wavenumber that enters the horizon at the first PBH-radiation equality, we find that $f_{\text{nl-cut}}/f_{\rm eva}\approx 10^{-5}(\Omega^{\rm peak}_{\rm GW,nl-cut})^{-3/7}$. Then, since we have that \begin{align}\label{eq:feva}
f_{\rm eva}\approx 0.7\,{\rm Hz}\left(\frac{M_{{\rm PBH},f}}{10^4{\rm g}}\right)^{-3/2}\,,
\end{align}
the position of $f_{\text{nl-cut}}$ is determined by the amplitude of the GW spectrum and the mass of the PBHs. }
\begin{align}\label{eq:GWsfnl-cut}
  \Omega^{\rm peak}_{\text{GW,nl-cut}} \approx 5\times10^{-11}\left(\frac{\beta}{10^{-5}}\frac{M_{\rm PBH,f}}{10^2\,{\rm g}}\right)^{-14/9}
   e^{-\left(\frac{{\cal A}^{\rm CMB}_{\calR}}{2\times 10^{-9}}\right)^{-1/2}\left(\frac{\beta}{10^{-5}}\frac{M_{\rm PBH,f}}{10^2\,{\rm g}}\right)^{-4/3}}\,.
\end{align}
The GW spectrum then quickly decays for $f>f_{\text{nl-cut}}$ and approximately goes as $f^7$ for $f<f_{\text{nl-cut}}$ until it smoothly transitions to an almost scale invariant plateau. The exponential dependence that appears in both the peak frequency \eqref{eq:fnl-cut} and amplitude \eqref{eq:GWsfnl-cut} comes from the logarithmic growth of matter fluctuations during the radiation dominated era \cite{Inomata:2020lmk}. We show the resulting GW spectrum in green in Fig.~\ref{Fig:plot}. For more details on the calculations we refer the reader to Ref.~\cite{Inomata:2020lmk}.\\

\noindent\textbf{\customlabel{iii}{\textit{(iii)}} GWs from PBH binaries. } 
The last source of GWs associated with PBHs that we consider are GWs from PBH binaries. Most of these PBH binaries form in the early universe (unless the PBH fraction as dark matter is very small around ${\rm f}_{\rm PBH}<10^{-15}$) and they do so via a three body interaction \cite{Nakamura:1997sm}. The two nearest PBHs fall towards each other but the third nearest PBH provides enough torque to the system to avoid a head-on collision. Then a very eccentric PBH binary is formed which eventually circularizes. For more details see the review \cite{Sasaki:2018dmp} and references therein. For recent advancements including the torque due to all PBHs, later interactions and accretion see Refs.~\cite{Raidal:2018bbj,Liu:2018ess,Liu:2019rnx,Vaskonen:2019jpv,Hutsi:2020sol,Wu:2020drm}. The merger rate at small redshift (in the nearby universe) per unit time and unit volume is given by \cite{Sasaki:2018dmp}\footnote{This estimate is valid if ${\rm f}_{\rm PBH}\ll1$. For large enough ${\rm f}_{\rm PBH}$, perhaps ${\rm f}_{\rm PBH}\sim 10^{-3}$, N-body interactions might become important.}
\begin{align}\label{eq:RR}
{\cal R}&\equiv \frac{dN_{\rm merge}}{dtdV}\approx  1.5\,\text{Mpc}^{-3} \text{yr}^{-1} \frac{{\rm f}_{\rm PBH}^2}{\left({\rm f}_{\rm PBH}^2+\sigma_{\rm eq}^2\right)^{21/74}}  \left(\frac{M_{\rm PBH}}{10^{-3}M_\odot}\right)^{-32/37}\,,
\end{align}
where $\sigma_{\rm eq}^2\approx 2.5\times 10^{-5}$. For instance, if ${\cal O}(10)$ solar mass PBHs explain some of the LVK events, which has a detected rate of around $10\,\,\text{Gpc}^{-3}\,\text{yr}^{-1}$, we need ${\rm f}_{\rm PBH}\sim 10^{-3}$.

For nearby PBH binaries one may be able to resolve the GW waveform. An approximate estimate for the maximum GW frequency before merging, in the source frame, is given by \cite{Maggiore:1900zz}
\begin{align}\label{eq:fmaxbinary}
f^{\rm max}_{\rm GW,binary}\approx 2.2\,{\rm kHz}\left(\frac{M_{\rm PBH,f}}{M_\odot}\right)^{-1}\,.
\end{align}
This frequency is twice the frequency associated with the innermost stable circular orbit. If the binary is at cosmological distances, the frequency today changes by a factor $1/(1+z)$ where $z$ is the redshift. However, if PBH binaries are too far or too weak to be resolved, they contribute to the GW background. For the calculation of the GW background from PBH binaries with a monochromatic PBH mass function see, \textit{e.g.}, Ref.~\cite{Wang:2019kaf} and references therein (for the energy spectrum of binary black holes during the whole inspiral-merger-ringdown phase in the non-spinning limit see Refs.~\cite{Ajith:2007kx,Ajith:2009bn}). For an example of a broad PBH mass function see Ref.~\cite{Braglia:2021wwa}. Here we use an analytical estimate of the peak of the GW spectrum from Ref.~\cite{Domenech:2021odz} for a monochromatic PBH mass function, which reads
\begin{align}\label{eq:GWbinaries}
\Omega^{\rm max}_{\rm GW,binary}h^2\approx 1.6\times 10^{-8}\left(\frac{M_{\rm PBH,f}}{M_\odot}\right)^{5/37}\left(\frac{{\rm f}_{\rm PBH}}{0.01}\right)^{53/37}\,.
\end{align}
The peak of the GW background from unresolved binaries comes from the nearest binaries and, therefore, the peak frequency is close to the maximum frequency of Eq.~\eqref{eq:fmaxbinary}. For $f>f^{\rm max}_{\rm GW,binary}$ the spectrum has a sharp cut-off and for $f<f^{\rm max}_{\rm GW,binary}$ the GW spectrum decays as $f^{2/3}$.  We show the GW spectral density in purple in Fig.~\ref{Fig:plot}. This completes our list of estimates for the GWs associated with PBHs.

\section{Testable PBH mass range at terrestrial GW detectors \label{sec:testPBH}}

With the estimates derived in \S~\ref{subsec:GWsignatures} we are ready to understand which PBH scenarios can be probed by terrestrial GW detectors. We note that our main aim is to recount the potential of terrestrial GW detectors to test PBH scenarios by providing order of magnitude estimates for future studies and searches. We will not dwell into details of how accurately future GW detectors may be able to probe PBH scenarios nor how well they may discern GWs from PBH scenarios from other GW sources. Instead, we refer the interested reader
to Refs.~\cite{Chen:2019irf,Miller:2020kmv,Bavera:2021wmw,Braglia:2022icu,Franciolini:2023opt}. As for terrestrial GW detectors, we consider a frequency range roughly from ${\rm Hz}$ to $10\,{\rm kHz}$. When needed,  we consider the peak sensitivity of Einstein Telescope which is most sensitive around $100\,{\rm Hz}$ with $\Omega_{\rm GW,0}h^2\sim 10^{-9}$. To be more optimistic, we also consider the power-law integrated sensitivity curve \cite{Thrane:2013oya} which gives a sensitivity to GW backgrounds of $\Omega_{\rm GW,0}h^2\sim 10^{-13}$ around $100\,{\rm Hz}$, after accumulating several years of data. In the case when the GW peak frequency is higher, we find the parameter space for which the low frequency tail enters a power-law integrated sensitivity curve.

\begin{figure}
\includegraphics[width=0.95\textwidth]{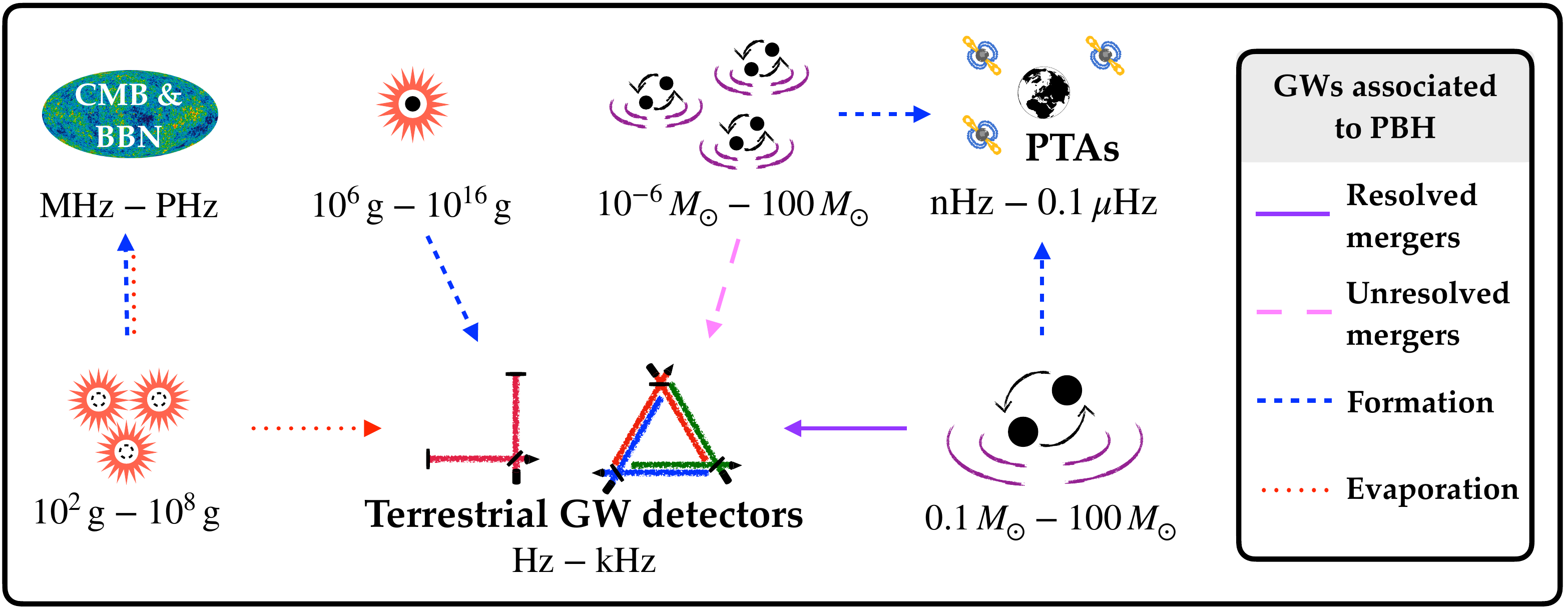}
\caption{Illustration of the PBH mass ranges that may be accessible to terrestrial GW detectors together with other complementary GW signals. Describing from left to right: First, we have PBHs evaporated before BBN ($M_{\rm PBH,f}\sim 10^{2}\,{\rm g}-10^{8}\,{\rm g}$) which can be detected from GWs associated with PBH reheating as well as high frequency GWs associated with formation and GWs from Hawking evaporation using CMB and BBN. Second, we have evaporated PBHs ($M_{\rm PBH,f}\sim 10^{6}\,{\rm g}-10^{16}\,{\rm g}$) that can be detected via GWs associated with PBH formation. Third and fourth, we respectively have GWs from unresolved ($M_{\rm PBH,f}\sim 10^{-6}\,M_\odot-100\,M_\odot$) and resolved ($M_{\rm PBH,f}\sim 0.1\,M_\odot-10^{2}\,M_\odot$) PBH binaries. Their GWs associated with PBH formation are low frequency GWs within the range of PTAs. Other PBH mass ranges are accessible to space-based GW detectors, such as LISA and Taiji, PTAs as well as high frequency GW detectors.\label{fig:diagram}}
\end{figure}

The main message of this and the next sections is summarized in Fig.~\ref{fig:diagram}. The estimates of the PBH mass ranges testable at terrestrial GW detectors are also summarized in more detail in Tab.~\ref{tab:table2}. At the end of this section, we also show an example of the GW spectral density together with sensitivity curves in Fig.~\ref{Fig:plot}. Now, we proceed to describe Fig.~\ref{fig:diagram} based on our discussions in \S~\ref{subsec:GWsignatures}. We start with GWs associated with evaporated PBHs and then turn to GWs associated with PBH binaries.

\begin{table}
\def\arraystretch{1.75}
\setlength\tabcolsep{1.5mm}
\begin{tabular}{|c|c|c|c|c|}
\toprule[2.0pt]\addlinespace[0mm]
\makecell{\textbf{GW background} \\\textit{associated with} PBH} & \makecell{\textbf{Formation}  \\ {(Adiabatic iGWs)}}  & \makecell{\textbf{Reheating I}   \\{(Isocurvature iGWs)}} & \makecell{\textbf{Reheating II}   \\{(Adiabatic iGWs)}} & \makecell{\textbf{Mergers}  \\{(Unresolved binaries)} }\\
\hline
\makecell{\makecell{\textbf{Mass range}  \\{(seen at $\rm Hz- kHz$)} }} &$10^6\,{\rm g}-10^{16}\,{\rm g}$& $10^3\,{\rm g}-10^{8}\,{\rm g}$& $10^2\,{\rm g}-10^{4}\,{\rm g}$ & $10^{-6}\,M_\odot-10^{2}\,M_\odot$\\
\toprule[2.0pt]
\end{tabular}
\caption{Summary of the mass ranges testable by terrestrial GWs detectors with the GW signals explained in points \ref{i}, \ref{ii} and \ref{iii}. Detailed explanations on how to derive such estimates is given in points  \ref{itemone}, \ref{itemtwo}, \ref{itemthree}, \ref{itemfour} and \ref{itemfive}. When needed, we take the power-law integrated sensitivity of ET, which is most sensitive at $100\,{\rm Hz}$ with $\Omega_{\rm GW,0}h^2\sim 10^{-13}$. \label{tab:table2}}
\end{table}

\subsection{Probing evaporated PBHs}

As evaporated PBHs cannot be detected directly, we only have the possibility to find their associated induced GW signal. This includes induced GWs associated with PBH formation and to PBH reheating, respectively discussed in points \ref{i} and \ref{ii}. Looking at their respective estimates for the frequency, Eqs.~\eqref{eq:ffomration}, \eqref{eq:fpossion} and \eqref{eq:fnl-cut}, we see that in order to enter the frequency range of terrestrial GW detectors, GWs associated with PBH formation require larger PBH masses than GWs associated with PBH reheating. Inputting some numbers in the estimates we find that:
\begin{enumerate}
    \item PBHs with masses between $10^8\,{\rm g}-10^{16}\,{\rm g}$ have the peak frequency \eqref{eq:ffomration} of GWs associated with PBH formation inside the frequency window. One may also be able to probe the amplitude the primordial spectrum almost down to ${\cal A}_{\calR}\sim 10^{-5}$ in \eqref{eq:lognormalpeak} (see Ref.~\cite{Romero-Rodriguez:2021aws} for a detailed analysis). \label{itemone}
    \item PBHs with masses between $10^6{\rm g}-10^{8}\,{\rm g}$ lead to GWs associated with PBH formation detectable only through its low frequency tail. However, one needs ${\cal A}_{\calR}> 10^{-2}$ in \eqref{eq:lognormalpeak}. \label{itemtwo}
    \item PBHs with masses between $10^3\,{\rm g}-10^{8}\,{\rm g}$ can be probed via isocurvature induced GWs associated with PBH reheating, since the peak frequency \eqref{eq:fpossion} is inside the observable frequency range. From Eqs.~\eqref{eq:fpossion} and \eqref{eq:GWspossion} we find that one could probe an initial PBH fraction from $\beta>6\times 10^{-9}$, where we used the peak sensitivity of the power-law integrated sensitivity curve of ET.
    \label{itemthree}
    \item PBHs with masses between $10^2\,{\rm g}-10^{4}\,{\rm g}$ yield an adiabatic induced GWs from PBH reheating accessible to terrestrial GW detectors. Such mass range comes from using Eq.~\eqref{eq:fnl-cut} and requiring the maximum amplitude possible in the GW spectrum \eqref{eq:GWsfnl-cut}.\footnote{This imposes ${\beta M_{\rm PBH,f}}\sim  10^{-2}\left(\tfrac{{\cal A}^{\rm CMB}_{\calR}}{2\times 10^{-9}}\right)^{-3/8}$.}
    And once the GW amplitude is fixed, the peak frequency \eqref{eq:fnl-cut} only depends on the PBH mass. We note though that relaxing the non-linear cut-off would enhance the amplitude of the GWs and broaden the parameter space.\label{itemfour}
\end{enumerate}

For Point \ref{itemone}, it is interesting to note that, although the fraction of PBHs in the mass range $10^9\,{\rm g}-10^{17}\,{\rm g}$ is tightly constrained \cite{Carr:2009jm,Poulter:2019ooo,Carr:2020gox,Romero-Rodriguez:2021aws}, one may still be able to probe the induced GWs associated with their formation. Points \ref{itemtwo}, \ref{itemthree} and \ref{itemfour} show that we will be able to probe the existence of very small PBHs in the early universe, although they evaporated well before BBN. For Points \ref{itemthree} and \ref{itemfour}, it would be intriguing to derive more accurate estimates using numerical simulations.

\subsection{Probing earth-to-solar mass PBHs}

Long-lived PBHs may be seen directly or indirectly by terrestrial GW detectors via GWs from PBH binaries, resolved or unresolved. The peak frequency of such GWs is given by Eq.~\eqref{eq:fmaxbinary}. We then classify two possibilities:
\begin{enumerate}
    \setcounter{enumi}{4}
    \item PBHs with masses between $0.1\,M_\odot-100\,M_\odot$ may be directly detected by terrestrial GW detectors. In fact, there is the possibility that LVK may have already detected PBHs \cite{Bird:2016dcv,Sasaki:2016jop,Wong:2020yig,Hutsi:2020sol,Franciolini:2021tla,Clesse:2020ghq,Morras:2023jvb,LIGOScientific:2022hai,Prunier:2023cyv}.\label{itemfive}
    \item PBHs with masses between $10^{-6}\,M_\odot-100\,M_\odot$ may yield a GW background signal from unresolved mergers. The mass range from $10^{-6}\,M_\odot-10^{-2}\,M_\odot$ can be probed with the low frequency tail of the GW spectrum \eqref{eq:GWbinaries}. In deriving this mass range we assumed ${\rm f}_{\rm PBH}\sim 10^{-2}$ which is consistent with current observations. A lower fraction of PBH as dark matter would yield a lower signal and a smaller mass range.\label{itemsix}
\end{enumerate}

We also note that it may be possible to test the mass range from $10^{-6}\,M_\odot-10^{-2}\,M_\odot$ via continuous GWs \cite{Miller:2020kmv,Pujolas:2021yaw} (see also Ref.~\cite{Alestas:2024ubs}), if the PBH fraction is not too small, around ${\rm f}_{\rm PBH}\sim {\cal O}(10^{-2})$. For the mass ranges where there is overlap with astrophysical black holes, namely for $M_{\rm PBH,f}>M_\odot$, one must carry out population analyses \cite{Franciolini:2021xbq,DeLuca:2021wjr,DeLuca:2021hde,Ng:2022agi,Franciolini:2023opt}, study the statistical nature of the GW background \cite{Braglia:2022icu}, or search for GW background anisotropies \cite{Wang:2021djr,Profumo:2023ybp}, in order to distinguish PBHs from astrophysical BHs. Interestingly, earth-to-solar mass PBHs may be tested by other complementary means such as microlensing or low frequency GWs, which we discuss in more detail in the next section.

\begin{figure}
\includegraphics[width=0.7\columnwidth]{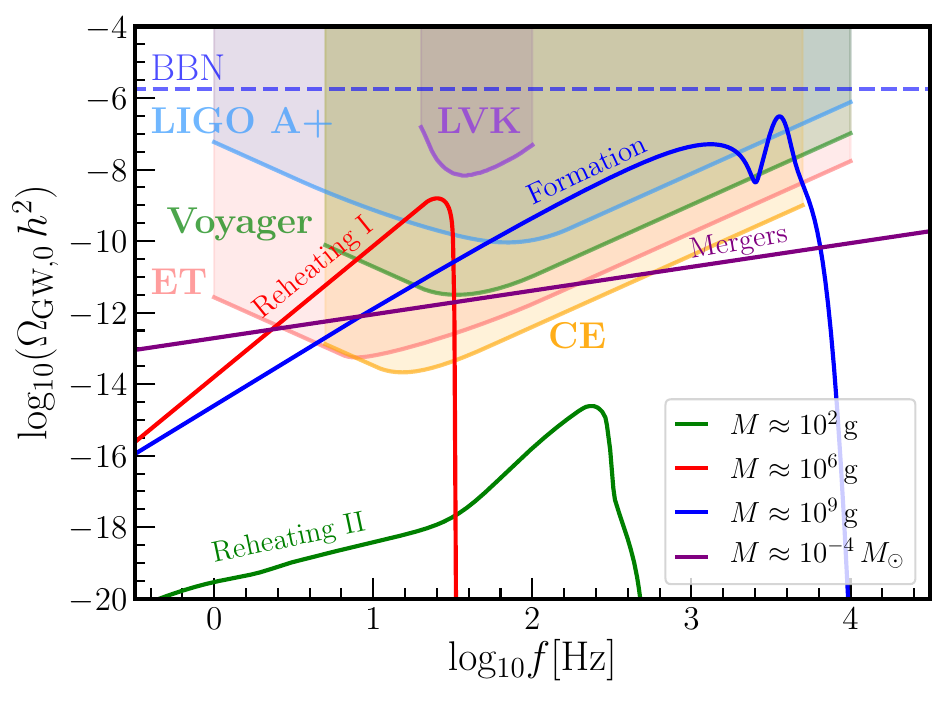}
\caption{Spectral density of GWs vs frequency in the range relevant for terrestrial GW detectors. We show one example for each GW signal associated with PBH discussed in this review (see points \ref{i}, \ref{ii} and \ref{iii}). In solid blue we show the GW spectrum of GWs associated with PBH formation. We considered $M_{\rm PBH,f}\approx 10^9\,{\rm g}$ and a log-normal primordial spectrum \eqref{eq:lognormalpeak} with $A_\calR=0.1$ and $\Delta=0.1$. We computed the GW spectrum using SIGWfast \cite{Witkowski:2022mtg}. In solid red we show the GW spectrum of isocurvature GWs associated with PBH reheating for $M_{\rm PBH,f}\approx 10^6\,{\rm g}$ and $\beta\approx2\times 10^{-8}$. In solid green we show the adiabatic GWs associated wwith PBH reheating for $M_{\rm PBH,f}\approx 10^2\,{\rm g}$ and $\beta\approx3\times 10^{-5}$, which was kindly provided by Keisuke Inomata. We believe that although the green line does not enter the observable window, the non-linear cut-off imposed by Ref.~\cite{Inomata:2020lmk} largely underestimates the GW signal. In solid purple we show the low frequency tail of the GW background from PBH binaries with $M_{\rm PBH,f}\approx 10^{-4}\,M_\odot$ and ${\rm f}_{\rm PBH}\approx 10^{-2}$ \cite{Wang:2019kaf}. We also show the power-law integrated sensitivity curves~\cite{Thrane:2013oya} for Einstein Telescope (ET), Cosmic Explorer (CE), Voyager and LIGO A+ experiments (see Refs.~\cite{ce,A+,voyager,Schmitz:2020syl} for the sensitivity curves). In light blue we plot the upper bounds from the LVK collaboration \cite{KAGRA:2021kbb}. The blue dashed line shows the current constraint from BBN \cite{Cyburt:2004yc,Arbey:2021ysg,Grohs:2023voo}.\label{Fig:plot}}
\end{figure}

\section{Complementary probes \label{sec:complementary}}

This review focused on the role of terrestrial GW detectors in testing PBH scenarios. However, there are other promising ways to test PBHs and complement the information from terrestrial GW detectors. For instance, we may use Big Bang Nucleosynthesis (BBN)
predictions and CMB observations, microlensing of electromagnetic waves and GW detectors in other frequency ranges. We list and describe them below.\\

\noindent\textbf{\customlabel{a}{\textit{(a)}} BBN \& CMB. } BBN predictions \cite{Cyburt:2004yc,Cooke:2013cba,Fields:2019pfx,Grohs:2023voo} as well as CMB observations \cite{Smith:2006nka,Clarke:2020bil} are sensitive to the presence of additional relativistic particles (sometimes also called dark radiation). Constraints from BBN and CMB are then usually parametrized with an effective number of additional relativistic species, denoted by $\Delta N_{\rm eff}$. Current limits from BBN \cite{Arbey:2021ysg} and CMB \cite{Planck:2018vyg} respectively give $\Delta N_{\rm eff}\lesssim 0.5$ and $\Delta N_{\rm eff}\lesssim 0.3$. Future CMB experiments, such as CMB-S4 \cite{CMB-S4:2016ple} might reach $\Delta N_{\rm eff}\lesssim 0.02$. The crucial point is that GWs with frequencies $f\gtrsim 10^{-10}\,{\rm Hz}$ and $f\gtrsim 10^{-15}\,{\rm Hz}$ may be considered as a dark radiation fluid, respectively, at the time of BBN and CMB \cite{Smith:2006nka}. 
Thus, BBN and CMB provide an integrated constraint on the total spectral density of GWs above these frequencies, which roughly yields $\Omega_{\rm GW,0}h^2\lesssim 10^{-6}$ \cite{Smith:2006nka,Clarke:2020bil} (see also Ref.~\cite{Caprini:2018mtu} sec.~4.1 for a summary with a nice explanation).

Regarding the GWs associated with PBHs, BBN and CMB open the possibility to test a considerable amount of high frequency GWs. This is relevant for the GWs associated with PBHs reheating for $1\,{\rm g}<M_{\rm PBH,f}<10^3\,{\rm g}$ as well as the GWs associated with PBH formation for $1\,{\rm g}<M_{\rm PBH,f}<10^6\,{\rm g}$, since they are not accessible to terrestrial GW detectors. Most interestingly, BBN and CMB might provide additional information on the PBH reheating scenario. GWs from Hawking evaporation of spinning PBHs is within the reach of future CMB-S4 experiments \cite{Hooper:2020evu,Masina:2020xhk,Arbey:2021ysg,Ireland:2023avg,Cheek:2022dbx} (see also Ref.~\cite{Ireland:2023zrd} for PBH evaporation with large extra dimensions). Such additional signatures might help in discerning the formation mechanism of PBHs for $M_{\rm PBH,f}>10^3\,{\rm g}$ \cite{Domenech:2021wkk}.\\

\noindent\textbf{\customlabel{b}{\textit{(b)}} Microlensing. } Electromagnetic waves, \textit{e.g.} from stars, travel through the dark matter halos and, if dark matter is composed by PBHs, one expects a certain amount of lensing events depending on ${\rm f}_{\rm PBH}$ \cite{Paczynski:1985jf} (see also sec.~3.1 of \cite{Sasaki:2018dmp} for a detailed explanation of microlensing). Current bounds from the absence of microlensing events set a constraint of about ${\rm f}_{\rm PBH}\lesssim{\cal O}(10^{-2})$ for a PBH mass range between $10^{-10}\,M_\odot-10\,M_\odot$ \cite{Carr:2020gox}.\footnote{For PBHs smaller than $10^{-10}\,M_\odot$ the Einstein radius becomes smaller than the size of the light source and strongly suppresses lensing \cite{Smyth:2019whb}.} However, most interesting are the microlensing candidate events reported by HSC \cite{Niikura:2017zjd} and OGLE \cite{Niikura:2019kqi} respectively with masses about $10^{-8}M_\odot$ and $10^{-4}M_\odot$. We note that $10^{-4}M_\odot$ PBHs might also be linked to the reported PTA signal \cite{Inomata:2023zup}. For another example, see Ref.~\cite{Sugiyama:2020roc} where a broad PBH mass function may explain the reported Pulsar Timing Array (PTA) signal as well as PBHs as dark matter, and can be tested by microlensing observations.\\

\noindent\textbf{\customlabel{c}{\textit{(c)}} PTAs. } Another complementary window to terrestrial GW detectors are nHz GWs which may be probed by Pulsar Timing Arrays (PTAs). From the estimate of the peak frequency of GWs associated with PBH formation, Eq.~\eqref{eq:ffomration}, we see such GWs fall in the PTA range for ${\cal O}(10 M_\odot)\gtrsim M_{\rm PBH,f}\gtrsim {\cal O}(10^{-3} M_\odot)$. If we consider the low frequency tail of the GW spectrum then the range might be extended down to $M_{\rm PBH,f}\gtrsim 10^{-9}M_\odot$ with future SKA sensitivity \cite{Janssen:2014dka}. Most interesting though are the current results from PTA data \cite{NANOGrav:2023gor,NANOGrav:2023hde,EPTA:2023fyk,EPTA:2023sfo,EPTA:2023xxk,Reardon:2023gzh,Zic:2023gta,Reardon:2023zen,Xu:2023wog,InternationalPulsarTimingArray:2023mzf}, which seems to suggest $M_{\rm PBHs}\sim 10^{-4}\,M_\odot-10^{-3}\,M_\odot $, if interpreted as an induced GW signal associated with PBH formation. Such a mass range is very interesting as it has implications for the microlensing events reported by OGLE and it may have a detectable GW background from unresolved PBH binaries, as pointed out by Ref.~\cite{Inomata:2023zup}. $\mu$Hz GW detectors like $\mu$-Ares \cite{Sesana:2019vho} (see also Refs.~\cite{Fedderke:2021kuy,Blas:2021mpc,Blas:2021mqw}) would extend the range of PTAs and provide more evidence for the PBH interpretation and extend the testable PBH mass range.\\

\noindent\textbf{\customlabel{d}{\textit{(d)}} Space-based GW detectors.} Future GW detectors such as LISA \cite{Barke:2014lsa,LISACosmologyWorkingGroup:2022jok}, Taiji \cite{Ruan:2018tsw}, TianQin \cite{Gong:2021gvw} and  DECIGO \cite{Yagi:2011wg,Kawamura:2020pcg}, will bridge PTAs and $\mu$Hz GW detectors with terrestrial GW detectors. This will offer the opportunity to test the low frequency tail of GW signals associated with PBHs that enter the terrestrial GW detector’s window as well as, of course, to extend the testable PBH mass ranges. Details on the capabilities of LISA to test PBH scenarios can be found in the review by the LISA cosmology working group \cite{LISACosmologyWorkingGroup:2023njw}. \\

\noindent\textbf{\customlabel{e}{\textit{(e)}} MHz-GHz GW detectors. } GWs associated with PBH formation and reheating of light PBHs as well as the mergers of planet-mass PBHs (see \textit{e.g.} Refs.~\cite{Herman:2020wao,Franciolini:2022htd} for detailed studies) are the sources of high frequency GWs. Even GWs from Hawking evaporation in the PBH reheating scenario might be testable by MHz-GHz GW detectors \cite{Ireland:2023avg,Ireland:2023zrd}. We note that although there are interesting events detected at current MHz GW detectors \cite{Goryachev:2021zzn} and the frequency could be explained by the merger of planet-mass PBHs, it seems an extremely unlikely explanation given the current sensitivity \cite{Domenech:2021odz} (see also Ref.~\cite{Lasky:2021naa}). Nevertheless, an improved sensitivity in future high frequency GW detectors will present an exciting window to further test PBH scenarios \cite{Aggarwal:2020olq,Berlin:2021txa,Bringmann:2023gba,Campbell:2023qbf}. And, they will complement any possible signals seen at terrestrial GW detectors.\\

\noindent\textbf{\customlabel{f}{\textit{(f)}} Lensing of GWs.} Another interesting probe to PBH scenarios using GWs is the recently proposed lensing of GWs \cite{Oguri:2022fir,GilChoi:2023qrz,Savastano:2023spl}. Lensed GWs might be sensitive to dark matter halos substructure due to frequency dependent wave optic effects, which could probe PBHs as a fraction dark matter in the $M_\odot-10^5M_\odot$ range \cite{Oguri:2022fir}. It would be interesting to investigate in which circumstances one might probe lighter PBH masses.\\

\section{Discussion and Conclusions \label{sec:conclusions}}

Cosmic events that produced PBHs shook the spacetime, resulting in ripples that we see today as GWs. Such induced GWs are also produced if PBHs reheat the universe. The first type of induced GWs, that we called GWs associated with PBH formation, can test the presence of PBHs with masses $10^{6}\,{\rm g}-10^{16}\,{\rm g}$ in current and future terrestrial GW detectors, such as LVK, Einstein Telescope and Cosmic Explorer. The second type of induced GWs, here referred to as GWs associated with PBH reheating, offer means to probe even lighter PBHs with masses from $10^2\,{\rm g}$ to $10^8\,{\rm g}$. Thus, terrestrial GW detectors have the potential to find GW signals associated with evaporated PBHs, otherwise unexplorable. 

Interestingly, we may also be able to find hints of black hole remnants as dark matter \cite{Domenech:2023mqk}, if PBH evaporation leaves Planck mass remnants behind as suggested by some quantum gravity theories \cite{Chen:2014jwq,Hossenfelder:2012jw,Vidotto:2018hww,Eichhorn:2022bgu,Platania:2023srt} (see also Refs.~\cite{Ashtekar:2005cj,Rovelli:2014cta,Bianchi:2018mml,Kazemian:2022ihc}). If so, there is a unique initial PBH mass, that is $M\sim 5\times10^5\,{\rm g}$, that can reheat the universe and its remnants be the dark matter,  with a sharp prediction for the frequency of induced GWs peaked at $100{\rm Hz}$ \cite{Domenech:2023mqk}.

In addition to evaporated PBHs, black hole binaries of long-lived PBHs, those with masses between $10^{-6}\,M_\odot-10^2\,M_\odot$, also produce GW signals at reach of terrestrial GW detectors. Most interesting for the PBH scenario is the possibility of finding evidence for planet-mass to sub-solar mass black holes, as their origin can only be primordial. And, there are potential candidates for such small PBHs in the LVK data \cite{Clesse:2020ghq,Morras:2023jvb,LIGOScientific:2022hai,Prunier:2023cyv}, for sub-solar mass PBHs, and in the microlensing data of OGLE \cite{Mroz:2017mvf,Niikura:2019kqi}, for earth-mass PBHs (HSC \cite{Niikura:2017zjd} also reported one candidate even from a moon-mass object). One may search for such small PBHs with continuous GWs \cite{Miller:2020kmv,Pujolas:2021yaw} and the GW background from unresolved binaries \cite{Wang:2019kaf,Braglia:2021wwa,Atal:2022zux}. The latter most often peaks at MHz-GHz frequencies, which may be accessible to high frequency GW detectors \cite{Aggarwal:2020olq,Berlin:2021txa,Bringmann:2023gba,Campbell:2023qbf}.

PBHs with mass between $10^{-6}\,M_\odot-10^2\,M_\odot$ also have a low frequency GW background signal associated with their formation, roughly at nHz. This is particularly interesting considering the recent PTA results \cite{NANOGrav:2023gor,NANOGrav:2023hde,EPTA:2023fyk,EPTA:2023sfo,EPTA:2023xxk,Reardon:2023gzh,Zic:2023gta,Reardon:2023zen,Xu:2023wog,InternationalPulsarTimingArray:2023mzf}. If interpreted as induced GWs associated with PBH formation, the corresponding mass range is around $10^{-4}\,M_\odot-10^{-3}\,M_\odot$ \cite{Inomata:2023zup}. Some of these PBHs might already be seen by OGLE \cite{Mroz:2017mvf,Niikura:2019kqi} and, if so, there should be a GW background signal from PBH binaries to be tested by future terrestrial GW detectors \cite{Inomata:2023zup}.

In a different direction, it would also interesting to fully explore how GW background anisotropies (see, \textit{e.g.}, Ref.~\cite{Contaldi:2016koz,Bartolo:2019oiq,LISACosmologyWorkingGroup:2022kbp}) might help in gathering evidence for the presence of PBHs \cite{Scelfo:2018sny,Wang:2021djr,Dimastrogiovanni:2022eir,Profumo:2023ybp} and, perhaps, even the presence of primordial non-Gaussianity \cite{Bartolo:2019zvb,Li:2023qua,Yu:2023jrs}. Moreover, as discussed in the context of MHz GW detectors by Ref.~\cite{Lasky:2021naa}, high frequency GWs, \textit{e.g.} from the mergers of planet-mass PBHs, might leave a GW memory trail at terrestrial GW detectors \cite{McNeill:2017uvq}.

We end this review by collecting the various tables and illustrations that we hope will be helpful to the interested reader. Relevant estimates for the frequency and amplitude of the GWs are summarized in Tab.~\ref{tab:table1}. In Tab.~\ref{tab:table2} we provide details on the testable PBH mass range corresponding to the various GW signals discussed in \S~\ref{subsec:GWsignatures}, points \ref{i}, \ref{ii}, \ref{iii}. We also illustrated the different possibilities and other complementary probes using GWs at other frequencies in Fig.~\ref{fig:diagram}. And in Fig.~\ref{Fig:plot} we display the GW spectral shapes of the GW signals associated with PBHs for several different scnarios.

\section*{Acknowledgments} 
We thank A.~Escriv{\`a}, F.~K{\"u}hnel and R.~Sheth for useful discussions and K.~Inomata for kindly providing the numerical data on the adiabatic induced GWs associated with PBH reheating. We also thank the participants and organizers of the workshop “Focus Week on Primordial Black Holes”, held on November 2023 at Kavli IPMU (Tokyo University), for the interesting discussions.
G.D. is supported by the DFG under the Emmy-Noether program grant no. DO 2574/1-1, project number 496592360.
This work is also supported in part by the JSPS KAKENHI grant No. 20H05853.

\bibliography{refgwscalar.bib} 

\end{document}